\documentclass[aps,pre,onecolumn,showpacs,a4paper]{revtex4}
\usepackage{graphicx} 
\usepackage{amsmath}
\usepackage{amssymb}
\usepackage{amscd}

\begin{document}

\title{Nonlinear oscillator with parametric colored noise:
 some analytical results}
\author{Kirone Mallick}
 \affiliation{Service de Physique Th\'eorique, Centre d'\'Etudes de Saclay,
 91191 Gif-sur-Yvette Cedex, France}
 \email{mallick@spht.saclay.cea.fr}
\author{Philippe Marcq}
 \affiliation{Institut de Recherche sur les Ph\'enom\`enes Hors \'Equilibre,
 Universit\'e de Provence,
 49 rue Joliot-Curie, BP 146, 13384 Marseille Cedex 13, France}
 \email{marcq@irphe.univ-mrs.fr}
\date{\today}

\begin{abstract}
    The  asymptotic behavior of  a nonlinear  oscillator subject  to a
    multiplicative Ornstein-Uhlenbeck noise is investigated.  When the
    dynamics is expressed in  terms of energy-angle coordinates, it is
    observed  that the angle  is a  fast variable  as compared  to the
    energy.  Thus, an effective stochastic dynamics for the energy can
    be derived if the angular  variable is averaged out.  However, the
    standard elimination  procedure, performed earlier  for a Gaussian
    white  noise,   fails  when  the   noise  is  colored  because  of
    correlations between the noise  and the fast angular variable.  We
    develop  here  a  specific  averaging scheme  that  retains  these
    correlations.   This  allows   us  to  calculate  the  probability
    distribution  function (P.D.F.) of  the system  and to  derive the
    behavior of physical observables in the long time limit.
\end{abstract}
\maketitle 

 \section{Introduction}

 The mechanical model of a particle trapped in a nonlinear
 confining  potential and 
 subject to  external  noise has been widely studied to illustrate
 the interplay between randomness and nonlinearity in 
 a classical dynamical system  \cite{anishchenko,vankampen,strato}.  
 Because of the  external noise, some parameters of
 the system fluctuate with time and the equation of motion
 becomes a stochastic differential equation with multiplicative
 ({\it i.e.},  parametric) noise. Models with multiplicative noise 
 can undergo  purely noise-induced phase transitions
  \cite{toral,vandenbroeck,lefever}  that can be realized 
  experimentally, for example,   in 
   nonlinear electronic circuits \cite{kabashima}  or  in
 systems with hydrodynamic instabilities \cite{fauve}. 
   Besides,   the presence of  multiplicative  noise 
  provides a microscopic model for   the anomalous diffusion
   of a particle in a fluctuating field of force 
  in the limit of  vanishing damping rate
 \cite{lind2,bouchaud}.

 When the external randomness is represented  by  a Gaussian 
 white noise, many analytical  results can be derived from the Fokker-Planck
 equation. For example,  in \cite{philkir1},  we have
 calculated the growth  exponents and the associated generalized
 diffusion constants of a nonlinear oscillator with random 
 frequency  in the limit of  vanishing damping rate. 
  The  key  feature of the method is to derive 
 an effective first order Langevin equation for the action variable
 by averaging  out   the fast angular motion. The  effective
 low dimensional problem can then be solved analytically. 
 However, this method fails for colored noise. 
  In fact,  when the noise is colored, a new time scale, the coherence
 time of the noise,  appears in the problem  and    straightforward
 averaging over the  fast  angular variable  
 leads to erroneous  results  because  correlation terms  between the noise
 and the fast variable are eliminated. We could  
  analyze   only qualitatively  
  \cite{philkir2}  through self-consistent scaling arguments 
 the long time behavior of  a  nonlinear oscillator subject to  colored
 parametric noise  and found that it is   radically different from that
 observed  with  white noise. 
 
  In this work, we  develop  a consistent adiabatic averaging procedure
  that allows us  to derive  analytical results for the general 
  nonlinear oscillator  subject to an Ornstein-Uhlenbeck
   parametric noise. We calculate,   in the long time limit, the 
   Probability  Distribution Function (P.D.F.)
   and derive explicit formulae for  the mean values  of physical observables
   such as the energy, the velocity-square, the amplitude-square.  
   The method  we describe  here 
  is a generalization of  the  averaging technique  introduced in 
    \cite{philkir3} for    the  classical  pendulum with
   fluctuating  frequency.

  The outline of this paper is as follows. In section 2, we 
  describe the model  under  study  and review 
  results that have been obtained in previous works~:  
   we write the dynamical equation in terms
  of energy and angle coordinates  and observe that a  separation
  between fast and slow variables occurs in the long time limit. We then  
    give  the analytical expression of the P.D.F.  
  and the scaling of physical observables    in the white noise case. 
 In section 3,  we   first  recall the   scaling behavior 
 observed when the noise is colored 
   and show explicitly that the standard averaging method
 fails in this case.  We then  develop  an 
 averaging method that allows us  to deal with 
  an Ornstein-Uhlenbeck  noise.
 This technique is based on a  recursive
 transformation of coordinates where 
 the noise itself is treated as a dynamical variable.
  An effective Langevin
 dynamics is  derived for the slow variable that leads to  
  an analytic expression    for  the P.D.F. and 
   allows us   to calculate  the 
  behavior   of physical observables  in the long time limit.
Our results are finally  extended to include the effect
 of  a small dissipation  in the system.

 \section{Review of earlier  results}
\label{sec:white:aa}

  In this section, we   review  previously derived   results that will
  be relevant for our  present work.

\subsection{The model}

We  consider a nonlinear oscillator of amplitude $x(t)$, trapped
 in  a  confining potential ${\mathcal U}(x)$  and subject to a multiplicative
 noise $\xi(t)$~:
 \begin{equation}
   \frac{\textrm{d}^2 }{\textrm{d} t^2}x(t) 
   = - \frac { \partial{\mathcal U}(x)}{\partial x}  + x(t)\, \xi(t) \,.
 \label{dynamique}
\end{equation}
 The statistical properties of the random function $\xi(t)$
 need not be specified  at this stage. 
 We restrict our analysis to the case where the potential
  ${\mathcal U}$ behaves as  an even polynomial in $x$ when 
 $ |x| \rightarrow \infty$.  A suitable rescaling
  of $x$ allows us to write
 \begin{equation}
     {\mathcal U}   \sim \frac{ x^{2n}}{2n}
  \,\, \hbox{ with } \,\, n \ge  2 \,. 
 \label{infU}
\end{equation}
As the amplitude $x(t)$ of the oscillator grows with time,
  the   behavior of ${\mathcal U}(x)$ for 
$ |x| \rightarrow \infty$  only is  relevant and Eq.(\ref{dynamique})
 reduces to
  \begin{equation}
   \frac{\textrm{d}^2 }{\textrm{d} t^2}x(t) 
 + x(t)^{2n-1}  = x(t) \, \xi(t)  \,.
 \label{dyn2}
\end{equation}

 \subsection{Energy-angle coordinates}

 An important  feature of Eq.(\ref{dyn2}) is that the deterministic system
 underlying it (obtained by setting $\xi \equiv  0$)   is {\it integrable}.
 The associated  energy and the angle variable  are given by 
 \cite{philkir1}
\begin{equation}
  E =  \frac{1}{2}\dot x^2 + \frac{1}{2n} x^{2n} \, \,\, ,
  \;\;\; \hbox{ and } \;\;\;   \phi  = 
 \frac{ \sqrt{n}} { (2n)^{1/2n} } \int_0^ { {x}/{E^{{1}/{2n}}} }
   \frac{{\textrm d}u}{\sqrt{ 1 -  \frac{u^{2n}}{2n}}}    \, , 
\label{defphi}
\end{equation}
 where the angle $\phi$ is defined modulo the
 oscillation period $4 K_n$, with  
\begin{equation}
   K_n   = \sqrt{n} \int_0^1  \frac{{\textrm d}u}{\sqrt{ 1 - u^{2n}}} \, .
\label{nperiod}
\end{equation}
 In terms of the energy-angle  coordinates  $(E, \phi)$,
 the original variables  $(x ,\dot x)$  are given by  
 \begin{eqnarray}
          x &=&   E^{1/{2n}} \, {\mathcal S}_n
 \left( \phi  \right) , \label{solnxv2}\\                 
     \dot x &=& (2n)^{\frac{n-1}{2n}} E^{1/2} \,
  {\mathcal S}_n'\left( \phi  \right) \, ,
\label{solnxv3}
 \end{eqnarray}
  where the hyperelliptic function
  \cite{abram,byrd}  ${\mathcal S}_n$ is defined as
   \begin{equation}
{\mathcal S}_n(\phi) = Y \,  \leftrightarrow 
  \phi  = 
     \frac{ \sqrt{n}} { (2n)^{1/2n} } \int_0^Y 
   \frac{{\textrm d}u}{\sqrt{ 1 -  \frac{u^{2n}}{2n}}}      \,    .
  \label{hyperelli}
\end{equation}
The function     ${\mathcal S}_n$ and its derivative 
 with respect to $\phi$, ${\mathcal S}_n'$, satisfy the following 
 relation   
\begin{equation}
  {\mathcal S}_n'(\phi) = \frac{ (2n)^{ \frac{1}{2n}}}{\sqrt{n}} \left(
      1 -  \frac{({\mathcal S}_n(\phi))^{2n}}{2n} \right)^{\frac{1}{2}} .
 \label{derivS}
 \end{equation}

  The presence of external noise spoils the integrability
 of the dynamical system (\ref{dyn2}) but does
 not preclude the use of $(E,\phi)$ instead of $(x,\dot{x})$  as
 coordinates   in phase space. 
 Introducing  an   auxiliary  variable $\Omega$  defined as 
 \begin{equation}
 \Omega =   (2n)^{ \frac{n+1}{2n} } \,  E^{\frac{n-1}{2n}}  \, ,
 \label{Omega}
\end{equation}
 equation~(\ref{dyn2}) is  written  as a system of two coupled
 stochastic differential equations  \cite{philkir1,philkir2}
   \begin{eqnarray}
     \dot \Omega  &=& (n -1) \, {\mathcal S}_n(\phi)
 {\mathcal S}_n'(\phi) \, \xi(t)  
   \,  ,      \label{evolomega} \\
 \dot\phi  &=& \frac{\Omega}{ (2n)^{\frac{1}{n}}}
 - \frac{{\mathcal S}_n(\phi)^2}{\Omega}  \, \xi(t)  \, .
    \label{evolphi}
   \end{eqnarray}
 This system is  rigorously 
 equivalent to the original  problem   (Eq.~\ref{dyn2}) and has been  
  derived without any hypothesis on 
 the  nature of the parametric perturbation   $\xi(t)$  which can even 
 be   a deterministic function  or   may  assume 
  arbitrary statistical properties.
 This  external perturbation  $\xi(t)$ continuously
  injects energy into the system.
  We have shown analytically  in \cite{philkir1,philkir2} that,
 when $\xi$ is a Gaussian white  noise or a dichotomous Poisson noise,
 the typical value of   $\Omega$   grows algebraically with time.
 We    also verified numerically that the same behavior is true for 
 an Ornstein-Uhlenbeck noise. Therefore, as seen from  Eq.~(\ref{evolphi}),  
  the phase $\phi$ is a fast variable  in all cases of interest.
 Assuming that,  in the long time limit,  
$\phi$ is uniformly distributed  over the 
 interval $[0, 4K_n]$ of a period, we obtain, after  averaging
 Eqs.~(\ref{solnxv2})~and~(\ref{solnxv3}) over the angle variable,
 the following equipartition relations
\begin{eqnarray}
\langle  E   \rangle   &=&  \frac{n+1}{2n} \, \langle \dot x^2   \rangle \, ,
\label{equipn2}  \\
\langle \dot x^2   \rangle   &=&   \langle  x^{2n}   \rangle  \, .
\label{equipnxv}
\end{eqnarray}
 These identities  are in agreement with numerical simulations \cite{philkir1}.

\subsection{Asymptotic formula for the P.D.F. in the white noise case}
\label{sec:white}

     We  recall here  the results obtained 
  in \cite{philkir1}  for the case where
  $\xi(t)$ is  a Gaussian white noise
 of zero mean value and amplitude ${\mathcal D}$:
\begin{eqnarray}
       \langle \xi(t)  \rangle &=&   0   \, ,\nonumber \\
   \langle \xi(t) \xi(t') \rangle  &=&  {\mathcal D} \, \delta( t - t') .
   \label{defbruitblanc}
 \end{eqnarray}
  Integrating  out  the angular variable $\phi$
  from the  Fokker-Planck equation for  the P.D.F. $P_t(\Omega, \phi)$
  associated  with the  system~(\ref{evolomega},\ref{evolphi}),
   an   averaged   Fokker-Planck 
 equation for the marginal distribution  $\tilde{P}_t(\Omega)$
 is derived  \cite{philkir1}  
 \begin{equation}
   \partial_t {\tilde P}  =  \frac{  {\tilde {\mathcal D} }}{2} 
\left(  \partial_{\Omega}^2 {\tilde P} - \frac{2}{n-1} \, \partial_{\Omega}
      \frac{{\tilde P}}{\Omega}  \right)  \, , \,\,\,\,\,\,\hbox{ with  }
 {\tilde {\mathcal D} } =  {\mathcal D} 
   \frac { (n-1)^2}{n+1}   \, (2n)^{\frac{2}{n} } \,
    \frac {\Gamma(\frac{3}{2n})}{\Gamma(\frac{1}{2n})} \,
    \frac{\Gamma(\frac{3n+1}{2n})}{\Gamma(\frac{3n+3}{2n})}  \, ,
\label{nFPmoy}
\end{equation} 
$\Gamma(.)$   being  the Euler Gamma function \cite{abram}. 
 
  We define,  for sake of    conciseness,  the    following two parameters
\begin{eqnarray}
    \mu_2 \equiv  \overline{ {\mathcal S}_n^2 } &=&
    (2n)^{ \frac{1}{n}} 
    \frac{ \Gamma\Big( \frac{3}{2n} \Big)\Gamma\Big( \frac{n+1}{2n} \Big)     }
       { \Gamma\Big( \frac{1}{2n} \Big)\Gamma\Big( \frac{n+3}{2n} \Big)   }  
   \label{defmu2}  \, , \\
    \mu_4   \equiv   \overline{ {\mathcal S}_n^4 } &=& 
 (2n)^{ \frac{2}{n}} 
  \frac{ \Gamma\Big( \frac{5}{2n} \Big)\Gamma\Big( \frac{n+1}{2n} \Big)     }
       { \Gamma\Big( \frac{1}{2n} \Big)\Gamma\Big( \frac{n+5}{2n} \Big)   }\, ,
\label{defmu4} 
 \end{eqnarray}
  where the  overline  denotes  the  average over an angular period,
   for example 
 $\,  \overline{ {\mathcal S}_n^2 } = \frac{1}{4K_n} \int_0^{4K_n}
  {\mathcal S}_n^2 (\phi) \rm{d}\phi \, .$ 
 The relations~(\ref{defmu2}) and~(\ref{defmu4})  are obtained 
 by making the change of variable  $u =  {\mathcal S}_n(\phi)$,
 using Eq.~(\ref{derivS}) and evaluating  the Eulerian integral
 of the first kind ({\it i.e.}, Beta function)
    thus obtained \cite{philkir1,abram}.

 Using  Eqs.~(\ref{nFPmoy}) and~(\ref{defmu2}), we deduce that 
 the  Fokker-Planck  equation~(\ref{nFPmoy})   corresponds
  to the following  effective Langevin  dynamics 
 for the slow variable~$\Omega$ 
\begin{equation}
   \dot\Omega = (2n)^{ \frac{1}{n}}   \frac{  (n-1) {\mathcal D}  }{n+3}
  \, \frac{\mu_2}{\Omega}  +   \frac{ (2n)^{ \frac{1}{2n}} \, (n-1) }
 {  \sqrt{n+3} }  
     \sqrt{\mu_2 }\,  {\eta}(t)  \, ,
\label{nLangeff} 
\end{equation}  
where  ${\eta}(t)  $  is an  effective Gaussian white noise 
 with amplitude ${\mathcal D}$.

  From  Eq.~(\ref{nLangeff}), it is clear  that the variable 
  $\Omega$   has normal diffusive behavior with time:
  $ \Omega \sim \left({\mathcal D}t\right)^{\frac{1}{2}} \, .$
 The averaged  Fokker-Planck equation (\ref{nFPmoy}) is
  exactly  solvable, leading to  
 the following expression for the energy P.D.F. 
 \begin{equation}
   {\tilde P}_t(E) =
  \frac{1}{  \Gamma \left(\frac{n + 1}{2 (n-1)}\right)} \,
\frac{n-1}{n E} \,
\left( \frac{(2n)^{\frac{n+1}{n}} \; E^{\frac{n-1}{n}}}
{2 {\tilde{\mathcal D}} t}\right)^{\frac{n+1}{2 (n-1)}}
\exp \left\{ - 
\frac{(2n)^{\frac{n+1}{n}} \; E^{\frac{n-1}{n}}}
{2 {\tilde{\mathcal D}}  t}
 \right\} \, .
\label{pdfmult}
\end{equation} 
 Thus,   the asymptotic time dependence  of 
  all moments of the energy, amplitude or velocity
 can be calculated analytically and 
 agree  with numerical simulations  \cite{philkir1}.
 In particular, using 
  Eqs.~(\ref{Omega}),~(\ref{equipn2}) and~(\ref{equipnxv})
  the  following scaling relations are obtained
\begin{eqnarray}
          E      &\sim&  \left({\mathcal D}t\right)^{\frac{n}{n-1}} , 
  \nonumber   \\ 
          x      &\sim&  \left({\mathcal D}t\right)^{\frac{1}{2(n-1)}} ,
  \nonumber    \\ 
         \dot x  &\sim&  \left({\mathcal D}t\right)^{ \frac{n}{2(n-1)} }  .
\label{scalingwhite}
\end{eqnarray}
 The physical observables grow algebraically with time, and the
 associated anomalous diffusion exponents  depend  only  on the behavior
 of the confining potential at infinity.

\section{Averaging method for colored noise  and calculation
 of the P.D.F.}

 We now consider  $\xi(t)$  to be 
  a   colored  Gaussian noise  with  correlation time  $\tau$ 
  that is   obtained  from
 the  Ornstein-Uhlenbeck  equation 
  \begin{equation} 
 \frac{{\textrm d} \xi(t)}{{\textrm d} t} = -\frac{1}{\tau} \xi(t) -  
\frac{1}{\tau} \eta(t)  \, , 
  \label{OU}
\end{equation}
 where $\eta(t)$ is  a Gaussian white noise
 of zero mean value and  of  amplitude ${\mathcal D}$. In the stationary limit,
when $t, t' \gg \tau$, we find:
\begin{equation}
       \langle \xi(t)  \rangle =   0  \;\; \mathrm{and} \;\;
     \langle \xi(t) \xi(t') \rangle  =   
\frac{\mathcal D}{2 \, \tau}   \, {\rm e}^{-|t - t'|/\tau} \,.
   \label{deftau}
 \end{equation} 

 In section \ref{sec:color}, we recall qualitative scaling results
 derived in \cite{philkir2} for colored noise and   show
 that a straightforward elimination of the angular variable $\phi$,
 along the lines described above for white noise, leads to 
  erroneous results. 
  We  then  elaborate   an  averaging scheme 
 that  allows   us to derive 
 the long time behavior of the nonlinear oscillator
 subject to multiplicative Ornstein-Uhlenbeck noise~: 
 in subsection \ref{sec:transformation},  we define a new set
 of variables on the three dimensional phase space $(\Omega, \phi, \xi)$ 
 that retains the relevant correlations between the noise and the fast
 angular variable.  In subsections \ref{sec:averagingFP} and
 \ref{sec:effectiveLangevin}  we  derive  the averaged
 Fokker-Planck equation and the associated  effective low-dimensional 
 Langevin system,  respectively. Analytical  results are obtained 
 in  \ref{sec:results} for the non-dissipative system and  extended 
 in \ref{sec:dissipation} for a small dissipation rate.

\subsection{Qualitative  behavior in the presence of  colored noise}
\label{sec:color}

 In \cite{philkir2}, we deduced from a self-consistent scaling Ansatz
 that, when the noise has a finite correlation time $\tau$,
 the variable $\Omega$ has a subdiffusive behavior with time
 and scales as 
\begin{equation}
   \Omega \sim \left({\mathcal D}t\right)^{\frac{1}{4}} \, .
\label{scalingomegacolor}
\end{equation}
  Using  this equation  and   
 Eqs.~(\ref{Omega}),~(\ref{equipn2}) and~(\ref{equipnxv}),  
  the  following scaling relations   are obtained 
\begin{eqnarray}
   E  &\sim& \left(\frac{ {\mathcal D}t} {2 \tau^2} \right)^{\frac{n}{2(n-1)}} , 
  \nonumber   \\
 x  &\sim&   \left(\frac{ {\mathcal D}t} {2 \tau^2} \right)^{\frac{1}{4(n-1)}} ,
   \nonumber    \\ 
 \dot x &\sim&  \left(\frac{{\mathcal D}t} {2 \tau^2}\right)^{\frac{n}{4(n-1)}} , 
\label{scalingcolor}
\end{eqnarray}
where the factor ${{\mathcal D}t}/{2 \tau^2}$ is found by dimensional analysis.
Thus, 
 the  anomalous diffusion exponents are halved
 when the noise is colored. 
   These  colored noise scalings
 are observed when $ t \to \infty$, even if 
 the correlation time $\tau$ is arbitrarily small.  More precisely,  the
 crossover between the white noise scalings~ (\ref{scalingwhite})
 and the colored noise scalings~(\ref{scalingcolor}) occurs when 
 the period $T$ of the underlying deterministic oscillator
 (which is a decreasing function of its amplitude) is of
 the order of $\tau$, {\it  i.e.},  for a   typical time 
 $t_c \sim ({\mathcal D}\tau^2)^{-1}$. When  $t \ll t_c$,
 the angular period of the system is much larger than 
 the correlation time of the noise, which  thus acts  as if it were
 white.  When  $t \gg  t_c$, the noise is highly
 correlated over a period and its effect is smeared out 
 leading to a slower diffusion.

 The  qualitative  scalings~(\ref{scalingcolor}) have  been obtained 
  by elementary  arguments  \cite{philkir1,philkir2}.   
 However, an   analytical calculation  of the P.D.F.,   
  that would yield quantitative  formulae for the 
  physical observables  in the long time limit,  
 has remained out of reach. Indeed, 
 the   standard averaging technique, 
  which was  successfully applied to   white noise, 
 fails  for the Ornstein-Uhlenbeck process, as we show in the following. 

 The random oscillator~(\ref{evolomega},\ref{evolphi})
 and  the Ornstein-Uhlenbeck  process~(\ref{OU})
 form  a three dimensional stochastic system  driven by a white noise
 $\eta(t)$. The  Fokker-Planck equation  for the joint P.D.F.
 $P_t(\Omega,\phi,\xi)$  is given by
\begin{equation}
 \frac{ \partial P_t}{\partial t} = -(n -1) \frac{\partial }{\partial \Omega}
  \Big( \, {\mathcal S}_n(\phi)
 {\mathcal S}_n'(\phi) \, \xi \,  P_t \Big) 
  - \frac{\partial }{\partial \phi} \left( 
 \Big(\frac{\Omega}{ (2n)^{\frac{1}{n}}}
 - \frac{{\mathcal S}_n(\phi)^2}{\Omega}  \, \xi \, \Big) P_t \right)
 + \frac{1}{\tau} \frac{ \partial \xi  P_t}{\partial \xi}
 +\frac{{\mathcal D}}{2 \, \tau^2}\frac{ \partial^2 P_t}{\partial \xi^2}\,  .
\label{FPcolor}
\end{equation}
We now average this  Fokker-Planck
 equation  over the angular variable $\phi$  assuming
 that  the probability measure  for $\phi$ is  uniform over  
 the   interval $[0,4K_n]$   when $t \to \infty$. 
 We use the fact that
 the average of the  derivative with respect to $\phi$  of any function is zero:
  \begin{equation}
 \overline{ \partial_{\phi}(\ldots)} = 0  \, .
 \label{moydphi}
\end{equation}  
This implies in particular that 
\begin{equation}
 \overline{{\mathcal S}_n(\phi)
 {\mathcal S}_n'(\phi)}  = \frac{1}{2}
   \overline{ \partial_{\phi}{\mathcal S}^2_n(\phi)} = 0 \, .
\end{equation}  
  Using these properties, we  obtain the evolution equation
  for   the marginal distribution  $\tilde{P}_t(\Omega,\xi)$  
\begin{equation}
 \frac{ \partial \tilde{P}_t}{\partial t} = 
 \frac{1}{\tau} \frac{ \partial \xi  \tilde{P}_t}{\partial \xi}
 +\frac{{\mathcal D}}{2 \, \tau^2}\frac{ \partial^2 \tilde{P}_t}
 {\partial \xi^2}\,  .
\label{FPavcolor}
\end{equation}
This phase-averaged Fokker-Planck equation 
 corresponds to the  following Langevin dynamics for the 
  variables $(\Omega,\xi)$
\begin{eqnarray}
     \dot \Omega  &=&  0 \,  \nonumber \\
      \dot \xi   &=& -\frac{1}{\tau} \xi  -   \frac{1}{\tau} \eta(t).
 \end{eqnarray}
 This result  predicts that   $\Omega$
 is not stochastic anymore and  is  conserved.  The integration 
  over the angular variable  averages  out the noise
 itself and leads to   conclusions
 that are   blatantly wrong.      The reason for
 this failure is the following: when the period of the 
  underlying deterministic oscillator is less 
 than    $\tau$, 
  the  angular variable $\phi$  becomes    fast as compared to  both the
 energy   $E$  and the noise $\xi$. 
    The straightforward    elimination of $\phi$   disregards
  the correlation  between  $\phi$  and $\xi$   and  eliminates the noise
   as well. 
 A correct   averaging scheme  that  takes  into  account the correlation 
 between $\phi$  and the noise $\xi$ will   be developed below.

\subsection{Transformation of the equations}
\label{sec:transformation}

     We shall define  recursively a new set of variables on the 
 $(\Omega, \phi, \xi)$  phase space that permits
 the adiabatic elimination of the angular variable
 without eliminating the relevant correlations between
  $\phi$  and the noise $\xi$. Defining   
\begin{equation}
        Y =  \frac{\Omega^2}{ (2n)^{\frac{1}{n}}}   \, , 
\label{defY}
\end{equation} 
 we deduce  the dynamics of $Y$ 
 from Eqs.~(\ref{evolomega})~and~(\ref{evolphi}) 
\begin{equation}
   \dot Y  =  (n -1) \, 
  \frac{ 2 {\mathcal S}_n(\phi){\mathcal S}_n'(\phi) \dot\phi}
{1  -  \frac{{\mathcal S}_n(\phi)^2 \xi}{Y}  }\,   \xi
  =   2  (n -1) {\mathcal S}_n(\phi){\mathcal S}_n'(\phi) \dot\phi \,  \xi
  +  \frac{2(n -1) {\mathcal S}_n^3(\phi){\mathcal S}_n'(\phi)\dot\phi \,\xi^2}
  {Y} +    {\mathcal O}\big(Y^{-3/2}\big)   \, ,  
\label{evolY}
\end{equation} 
 we have neglected    here  terms  smaller than $Y^{-1}$
 ({\it i.e.},   of order strictly higher than 
  $Y^{-1}$).  The terms  
 ${\mathcal S}_n(\phi){\mathcal S}_n'(\phi) \dot\phi$ and
  ${\mathcal S}_n^3(\phi){\mathcal S}_n'(\phi) \dot\phi$  being exact 
   derivatives  with respect to time, we  integrate
  Eq.~(\ref{evolY})  by parts   and use Eq.~(\ref{OU})  to obtain
\begin{equation}
\frac{1}{n-1}  \dot Y  =  \frac{ \rm{d}}{\rm{d}t} 
  \Big( {\mathcal S}_n^2(\phi)  \xi    \Big)
  + {\mathcal S}_n^2(\phi) \frac{ \xi + \eta}{\tau} +
    \frac{ \rm{d}}{\rm{d}t}  
 \Big(  \frac{  {\mathcal S}_n^4(\phi)  \xi^2 }{ 2 Y}    \Big)
 + {\mathcal S}_n^4(\phi)\xi  \frac{ \xi + \eta}{\tau} 
+    {\mathcal O}\big(Y^{-3/2}\big)   \, . 
 \label{evol2Y}
\end{equation} 
  Collecting all time derivatives on the left hand side
     (l.h.s.),     we rewrite Eq.~(\ref{evol2Y}) as 
\begin{equation}
\frac{ \rm{d}}{\rm{d}t}   \Big(\frac{1}{n-1}    Y
   -  {\mathcal S}_n^2(\phi)  \xi - 
 \frac{  {\mathcal S}_n^4(\phi)  \xi^2 }{ 2 Y}  \Big)  =
   \mu_2  \frac{ \xi }{\tau} + \Big( {\mathcal S}_n^2(\phi) - \mu_2  \Big)
  \frac{ \xi }{\tau} + 
{\mathcal S}_n^2(\phi) \frac{\eta}{\tau}
 + {\mathcal S}_n^4(\phi)\xi  \frac{ \xi + \eta}{\tau} 
+    {\mathcal O}\big(Y^{-3/2}\big)   \, ,  
 \label{evol3Y}
\end{equation} 
 where $\mu_2$   is  defined in  Eq.~(\ref{defmu2}). 
  The function $({\mathcal S}_n^2(\phi) - \mu_2)$ is periodic in $\phi$
 and its   angular average
  vanishes identically  by virtue of  Eq.~(\ref{defmu2}). 
 However, the product of  this term  with  the  noise $\xi$, that
 appears in Eq.~(\ref{evol3Y}), does not average to zero (because of
 correlations between the angle variable and the noise). 
 Therefore, defining   a  $\phi$-periodic  function ${\mathcal D}_n(\phi)$
 that  satisfies
\begin{equation}  
 \frac{\textrm{d}\; {\mathcal D}_n(\phi)   }{\textrm{d} \phi}
 = {\mathcal S}_n^2(\phi)  - \mu_2 
\,\,\, \hbox{ and } \,\,\, \overline{ {\mathcal D}_n(\phi)}  = 0 \, , 
 \label{defDn}
\end{equation}
 we  rewrite the product $({\mathcal S}_n^2(\phi) - \mu_2) \xi$
 as follows  
 \begin{equation}  
   ({\mathcal S}_n^2(\phi) - \mu_2) \xi = 
\frac{\textrm{d}}{\textrm{d} t}  \Big( 
  \frac{ {\mathcal D}_n(\phi)  \xi }{(2n)^{-\frac{1}{2n}}  Y^{1/2}   }   \Big)
  + \frac{{\mathcal D}_n(\phi)  (\xi + \eta) }
 {  \tau  (2n)^{-\frac{1}{2n}} Y^{1/2}   }  + 
 \frac{{\mathcal S}_n^4(\phi) - \mu_2{\mathcal S}_n^2(\phi)  
 +  (n-1) {\mathcal D}_n(\phi) {\mathcal S}_n(\phi){\mathcal S}_n'(\phi)}
    {Y} \,  \xi^2  +    {\mathcal O}\big(Y^{-3/2}\big)    \, . 
\label{Ipp2}
\end{equation}
 This relation can be verified  by calculating the  time derivative 
 that appears   on the
  right hand side (r.h.s.)   
  and neglecting all terms  of the order  ${\mathcal O}\big(Y^{-3/2}\big)$. 
 Substituting Eq.~(\ref{Ipp2}) in Eq.~(\ref{evol3Y}) we finally
 obtain
\begin{eqnarray}
\frac{ \rm{d}}{\rm{d}t}   \Big(\frac{1}{n-1}   Y
   -  {\mathcal S}_n^2(\phi)  \xi 
 -  \frac{ {\mathcal D}_n(\phi)  \xi }{ \tau (2n)^{-\frac{1}{2n}} Y^{1/2}} 
 -  \frac{  {\mathcal S}_n^4(\phi)  \xi^2 }{ 2 Y}  \Big) 
 \nonumber &=& \\   
   \mu_2  \frac{ \xi}{\tau}  +  
\frac{{\mathcal D}_n(\phi)  \xi  }
 {  \tau^2  (2n)^{-\frac{1}{2n}} Y^{1/2}   } 
   &+& \frac{ 2 {\mathcal S}_n^4(\phi) - \mu_2{\mathcal S}_n^2(\phi)  
 +  (n-1) {\mathcal D}_n(\phi) {\mathcal S}_n(\phi){\mathcal S}_n'(\phi)}
    {\tau Y} \,  \xi^2   \nonumber \\   
 &+& \Big( {\mathcal S}_n^2(\phi)   +   \frac{{\mathcal D}_n(\phi)}
 {  \tau (2n)^{-\frac{1}{2n}} Y^{1/2}   } 
  +  \frac{  {\mathcal S}_n^4(\phi)  \xi }{  Y}\Big) \frac{\eta}{\tau}  \, .  
 \label{evol4Y}
\end{eqnarray}
  Considering the l.h.s. of this equation, it is natural to define
 a new dynamical variable $Z$ such that
\begin{equation}
    Z =  Y - (n-1) \Big(  {\mathcal S}_n^2(\phi)  \xi 
 +  \frac{ {\mathcal D}_n(\phi)  \xi }{ \tau (2n)^{-\frac{1}{2n}} Y^{1/2}} 
 +   \frac{  {\mathcal S}_n^4(\phi)  \xi^2 }{ 2 Y}   \Big) \, .
\label{defZ}
\end{equation}
 The two variables $Y$ and $Z$ are identical at leading order. 
 The dynamical
  equation for $Z$ is obtained   by writing   $Y$ in terms of 
 of $Z$  in   Eq.~(\ref{evol4Y}) and 
 neglecting all terms  of  order 
  strictly  higher  than  $Z^{-1}$.
 This change of variable is straightforward
  on the l.h.s. of   Eq.~(\ref{evol4Y}) and, because   $Y$ appears
 only   in the denominator  on the r.h.s.,   it is legitimate   
to  replace  $Y$ by  $Z$   (at  order $Z^{-1}$).  
  Finally, we obtain  
the following  system of coupled Langevin equations 
    \begin{eqnarray}
  \frac{1}{n -1}  \dot Z  &=&  {\mathcal J}_Z(Z,\phi,\xi)
 +  {\mathcal D}_Z(Z,\phi,\xi) \frac{ \eta(t)  }{\tau}
\label{evZ}   \,  ,   \\    
   \dot\phi  &=&   {\mathcal J}_{\phi}(Z,\phi,\xi) \, , 
    \label{evphi} \\   
    \dot \xi    &=&  -\frac{1}{\tau} \xi  -  \frac{1}{\tau} \eta(t)   \, ,
  \label{evxi} 
   \end{eqnarray}
where the current  and diffusion functions  are given by
   \begin{eqnarray} 
 { J}_Z(Z,\phi,\xi) &=& 
\mu_2  \frac{ \xi}{\tau}  +  
\frac{{\mathcal D}_n(\phi)  \xi  }
 {  \tau^2  (2n)^{-\frac{1}{2n}} Z^{1/2}   } 
   +  \frac{ 2 {\mathcal S}_n^4(\phi) - \mu_2{\mathcal S}_n^2(\phi)  
 +  (n-1) {\mathcal D}_n(\phi) {\mathcal S}_n(\phi){\mathcal S}_n'(\phi)}
    {\tau Z} \,  \xi^2   \,  ,  \label{courZ}  \\    
{\mathcal D}_Z(Z,\phi,\xi)  &=&  {\mathcal S}_n^2(\phi)   +   \frac{{\mathcal D}_n(\phi)}
 {  \tau (2n)^{-\frac{1}{2n}} Z^{1/2}   } 
  +  \frac{  {\mathcal S}_n^4(\phi)  \xi }{ Z}
   \,  ,  \label{diffZ} \\    
{ J}_{\phi}(Z,\phi,\xi)  &=&  
 \frac{\Omega(Z,\phi,\xi)}{ (2n)^{\frac{1}{n}}} 
 - \frac{{\mathcal S}_n^2(\phi)}{\Omega(Z,\phi,\xi)}  \, \xi     \, .
\end{eqnarray}
 We now write the Fokker-Planck equation  associated with this
 stochastic system and average it over the rapid variations
 of the angular variable $\phi$.

\subsection{Averaging the Fokker-Planck equation}
\label{sec:averagingFP}

 The Fokker-Planck equation  for the P.D.F.  $\Pi_t(Z,\phi,\xi)$ 
 corresponding to  the system~(\ref{evZ},\ref{evphi} and  \ref{evxi})  
 is given by 
\begin{eqnarray}
 \frac{ \partial \Pi_t}{\partial t} &=&
  - (n -1) \frac{\partial }{\partial Z}
  \left( { J}_Z    \,  \Pi_t \right) 
  - \frac{\partial }{\partial \phi}
  \left(  { J}_{\phi}  \,  \Pi_t \right)  
 + \frac{1}{\tau} \frac{ \partial \xi  \Pi_t}{\partial \xi} +  \nonumber \\
  &&  \frac{{\mathcal D}}{2 \, \tau^2} \Bigg\{
 (n-1)^2 
\frac{\partial }{\partial Z} {\mathcal D}_Z \frac{\partial }{\partial Z}
  ({\mathcal D}_Z  \Pi_t) -  (n-1) \frac{\partial^2 }{\partial Z\partial \xi   }
  ({\mathcal D}_Z  \Pi_t)  -  (n-1)   \frac{\partial }{\partial Z}
 {\mathcal D}_Z \frac{ \partial  \Pi_t}{\partial \xi} + 
  \frac{ \partial^2 \Pi_t}{\partial \xi^2}   \Bigg\}\,   .
\label{FPZ}
\end{eqnarray}
  We  now  integrate  out  the fast angular variable $\phi$
 from  this   equation, {\it i.e.}, we consider that
 in the long time  limit, the P.D.F.  $\Pi_t(Z,\phi,\xi)$ 
 becomes uniform in $\phi$ and reduces to   the function 
 $\tilde \Pi_t(Z, \xi)$  of the two variables $Z, \xi$. 
 Because  the   stochastic system~(\ref{evZ}, \ref{evphi} and  \ref{evxi})
 is valid up to  the order  ${\mathcal O}\big(Z^{-1}\big)$,   
   we should retain, while averaging Eq.~(\ref{FPZ}),   only 
  the terms  that scale at most as $Z^{-2}$
 (we recall that the derivative operator   $\frac{\partial }{\partial Z}$
 itself  scales  as $Z^{-1}$). 

 We now calculate the angular average
 of each term   appearing  on the r.h.s. of Eq.~(\ref{FPZ}). 
 To deal with the first term, we need to calculate the average
 of the current function  defined in Eq.~(\ref{courZ})   and  we obtain
 \begin{equation}
\overline{J_Z}(Z,\xi) =   
\mu_2  \frac{ \xi}{\tau}  +  
\frac{\overline{ {\mathcal D}_n(\phi)}  \xi  }
 {  \tau^2  (2n)^{-\frac{1}{2n}} Z^{1/2}   } 
   +  \frac{   2 \overline{ {\mathcal S}_n^4(\phi)} - \mu_2
     \overline{  {\mathcal S}_n^2(\phi) } 
 +  (n-1)\overline{ 
   {\mathcal D}_n(\phi) {\mathcal S}_n(\phi){\mathcal S}_n'(\phi)   }  }
    {\tau Z} \,  \xi^2   \, .
\end{equation}
 Integrating by parts  and  using  Eq.~(\ref{defDn}), we write 
 \begin{equation}
  \overline{ 
   {\mathcal D}_n(\phi) {\mathcal S}_n(\phi){\mathcal S}_n'(\phi)   }
  = -  \frac{1}{2}
 \overline{ {\mathcal S}_n^2(\phi) \Big({\mathcal S}_n^2(\phi)-\mu_2\Big)} 
   =  \frac{ \mu_2^2 - \mu_4 }{2}   \,   ,   
\label{eq:Ipp}
\end{equation}
where $\mu_4$ was defined in Eq.~(\ref{defmu4}).
Using Eqs.~(\ref{defmu2}),~(\ref{defmu4})~(\ref{defDn})~and~(\ref{eq:Ipp}), 
 $\overline{J_Z}(Z,\xi)$ is written as 
  \begin{equation}
\overline{J_Z}(Z,\xi) =  \mu_2  \frac{ \xi}{\tau}  +  
 + \frac{ \xi^2}   {\tau Z} \Big( 2\mu_4 - \mu_2^2 
  +\frac{n-1}{2}(\mu_2^2 - \mu_4 )   \Big)      \, .
 \label{eq:moyJZ}
 \end{equation}
  The average of
 the diffusion function  defined  in  Eq.~(\ref{diffZ})
 can also be calculated in a similar manner~: 
 \begin{equation} 
\overline{{\mathcal D}_Z}(Z,\xi) =  \mu_2 + \frac{\mu_4  \xi} { Z}  \, .
\label{eq:moyDZ}
 \end{equation}
 From this equation the angular averages of the second derivative
 terms  that appear  in Eq.~(\ref{FPZ}) are readily found
\begin{eqnarray}
\overline{ \frac{\partial }{\partial Z} 
  {\mathcal D}_Z \frac{\partial  }{\partial Z} ({\mathcal D}_Z \Pi_t ) }
&=&  \mu_4 \frac{\partial^2  \tilde\Pi_t    }{\partial Z^2}
\label{eq:moydZdZ} \, ,  \\
  \overline{ \frac{\partial^2 }{\partial Z\partial \xi }  ({\mathcal D}_Z  \Pi_t)} 
  &=&      \frac{\partial^2  }
 {\partial Z\partial \xi   } \Big(
    (  \mu_2 + \frac{\mu_4  \xi} { Z})   \tilde\Pi_t \Big)     \, ,
 \label{eq:moy1dZdxi}  \\
\overline{   \frac{\partial }{\partial Z}
 {\mathcal D}_Z \frac{ \partial  \Pi_t}{\partial \xi} }  &=&   
   \frac{\partial }{\partial Z} \Big( 
   (  \mu_2 + \frac{\mu_4  \xi} { Z}) 
  \frac{ \partial \tilde\Pi_t}{\partial \xi} \Big)    \, . 
\label{eq:moy2dZdxi}
 \end{eqnarray}
 In Eq.~(\ref{eq:moydZdZ}) we have retained only the leading 
 order  in the average of the square of
  the diffusion function  ${\mathcal D}_Z$,
 because the second derivative with respect to $Z$ already
 scales as $Z^{-2}$.

Finally, we deduce  from  Eq.~(\ref{moydphi}) that 
\begin{equation}
\overline{ \frac{\partial }{\partial \phi} (J_\phi \Pi_t) } = 0 \, .
 \label{eq:moydJphi}
\end{equation}
  
 Inserting the expressions obtained in 
 Eqs.~(\ref{eq:moydJphi}),
  (\ref{eq:moyJZ}),~(\ref{eq:moydZdZ}),~(\ref{eq:moy1dZdxi})
  and~(\ref{eq:moy2dZdxi}) in Eq.~(\ref{FPZ}), 
  we complete the derivation of the averaged Fokker-Planck equation~:
\begin{eqnarray}
 \frac{ \partial \tilde\Pi_t}{\partial t} =&&
  - (n -1) \frac{\partial }{\partial Z}
  \Bigg( \,\left(      \frac{\mu_2}{\tau} \xi 
 + \frac{ (n-3)\mu_2^2 -  (n-5) \mu_4 }   {2\tau Z} \xi^2   \right) 
     \tilde\Pi_t \Bigg) 
  + \frac{1}{\tau} \frac{ \partial \xi  \tilde\Pi_t}{\partial \xi}
  +  \nonumber \\
  &&  \frac{{\mathcal D}}{2 \, \tau^2}  \Bigg\{
 (n-1)^2  \mu_4 \frac{\partial^2  \tilde\Pi_t    }{\partial Z^2}
  -  (n-1) \frac{\partial^2  }
 {\partial Z\partial \xi   } \Big(
    (  \mu_2 + \frac{\mu_4 } { Z} \xi )   \tilde\Pi_t \Big)  
-  (n-1) \frac{\partial }{\partial Z} \Big( 
   (  \mu_2 + \frac{\mu_4 } { Z} \xi) 
  \frac{ \partial \tilde\Pi_t}{\partial \xi} \Big) 
+  \frac{ \partial^2 \tilde\Pi_t}{\partial \xi^2}   \Bigg\}\,   .
\label{eq:avFPZ}
\end{eqnarray}
 This  equation describes  a stochastic motion 
 in the  {\it two-dimensional} phase space $(Z,\xi)$. We shall now write
 the  effective  Langevin equations for  $(Z,\xi)$  that correspond to this 
  averaged Fokker-Planck equation.

\subsection{Effective Langevin equations}
\label{sec:effectiveLangevin}

 Consider the  following stochastic system
\begin{eqnarray}
  \frac{1}{n-1}\dot Z  &=&    \frac{\mu_2}{\tau} \xi +
   \frac{ (n-3)\mu_2^2 -  (n-5) \mu_4 }   {2\tau Z} \xi^2   
 + \frac{ \sqrt{\mu_4 - \mu_2^2}   }{\tau} \eta_1(t)
  +  \left(  \mu_2 + \frac{\mu_4 } { Z} \xi  \right) \frac{1}{\tau} \eta_2(t)  \, , 
  \label{eq:avZ}  \\
 \dot \xi    &=&  -\frac{1}{\tau} \xi  -  \frac{1}{\tau} \eta_2(t)   \, ,  
  \label{eq:avxi} 
\end{eqnarray}
 where $\eta_1(t)$ and $\eta_2(t)$ are two independent Gaussian
 white noises of amplitude ${\mathcal D}$. 
  The Fokker-Planck equation  associated with 
 the stochastic system~(\ref{eq:avZ}, \ref{eq:avxi})
  is identical  to   Eq.~(\ref{eq:avFPZ})    but   for the
 second derivative in $Z$. In Eq.~(\ref{eq:avFPZ})
 this term is given by
 \begin{equation}
  \frac{{\mathcal D}}{2 \, \tau^2}   
 (n-1)^2  \mu_4 \frac{\partial^2  \tilde\Pi_t    }{\partial Z^2}  \, , 
\end{equation}
  whereas the corresponding term for the  Fokker-Planck equation  
 corresponding to Eqs.~(\ref{eq:avZ}, \ref{eq:avxi}) is 
 \begin{equation}
  \frac{{\mathcal D}}{2 \, \tau^2}   
 (n-1)^2  \Bigg\{  (\mu_4 - \mu_2^2) \frac{\partial^2  \tilde\Pi_t    }{\partial Z^2}
 + \frac{\partial }{\partial Z} \left(   \mu_2 + \frac{\mu_4 } { Z} \xi \right) 
   \frac{\partial }{\partial Z} \left(   \mu_2 + \frac{\mu_4 } { Z} \xi \right)\tilde\Pi_t 
   \Bigg\} \, . \label{termdiff}
 \end{equation}
 Retaining  in Eq.~(\ref{termdiff})
  only contributions of order up to $Z^{-2}$, as done earlier,
 we observe that the two expressions coincide. 
  Hence,    the  Fokker-Planck equation corresponding to
 the system~(\ref{eq:avZ},\ref{eq:avxi})  is   
 precisely  given by the averaged  Fokker-Planck equation~(\ref{eq:avFPZ}) in the
 two-dimensional phase space $(Z, \xi)$ obtained after adiabatic
 elimination of the fast angular variable. This Langevin dynamics for the slow
 variable $Z$ plays a role similar to that of Eq.~(\ref{nLangeff}) in the case of
 white noise. However, the stochastic system~(\ref{eq:avZ},\ref{eq:avxi})
 does not admit  an exact solution and we  need to make further simplifications
 in order to obtain analytic results. The main idea is  to eliminate  the  
  white noise $\eta_2(t)$ from Eq.~(\ref{eq:avZ})  so that 
 the equation for $\dot Z$  is partially   decoupled 
 from that for $\dot \xi$. We therefore
  define a  new variable  $Z_2$ such that
  \begin{equation}
    Z_2  = Z + (n -1)\mu_2\xi +    (n -1) \frac{ \mu_4} { 2  Z} \,   \xi^2   \, .
  \label{eq:defZ2}
  \end{equation}
 The time evolution of $Z_2$ is given by 
\begin{equation}
   \dot   Z_2  =  \frac{(n-1)(3-n) (\mu_4 - \mu_2^2)}{ 2 \tau Z_2} \,   \xi^2
  + (n-1) \frac{ \sqrt{\mu_4 - \mu_2^2}   }{\tau} \eta_1(t)  \, .               
\label{eq:dynZ2}
  \end{equation}
 To derive this equation we retained only the  terms  that scale at most as 
 $Z_2^{ -1}$. Using the fact  that the mean value $\langle \xi^2 \rangle$
  of the Ornstein-Uhlenbeck process  is  
 equal to ${\mathcal D}/2\tau$,   we  write Eq.~(\ref{eq:dynZ2}) as follows
 \begin{equation}
   \dot   Z_2  =   \frac{(n-1)(3-n) {\mathcal D}}{ 4 \tau^2} \frac{\mu_4 - \mu_2^2}{Z_2}
  + (n-1)  \frac{ \sqrt{\mu_4 - \mu_2^2}   }{\tau} \eta_1(t)  
   +  \frac{(n-1)(3-n) (\mu_4 - \mu_2^2)}{ 2 \tau Z_2} \,  (\xi^2 - \langle \xi^2 \rangle) 
       \, .            \label{eq:evolZ2}
  \end{equation}
 This equation contains two random noise sources: the white noise $\eta_1(t),$
 and the nonlinear colored noise $\xi^2 - \langle \xi^2 \rangle$, of zero mean 
 and of  finite variance. This colored noise is multiplied by a prefactor
 $1/ Z_2$  and,  therefore, 
    its contribution   in the long time limit  becomes
   negligible as compared to that of the white noise. We thus discard this term
 and  obtain  the following white noise effective Langevin dynamics for 
 $Z_2$
\begin{equation}
   \dot   Z_2  =   \frac{(n-1)(3-n) {\mathcal D}}{ 4 \tau^2} \frac{\mu_4 - \mu_2^2}{Z_2}
  + (n-1)  \frac{ \sqrt{\mu_4 - \mu_2^2}   }{\tau} \eta_1(t)   \, .          
   \label{eq:evolZ2blanc}
  \end{equation}
 This equation has  the same mathematical structure  as Eq.~(\ref{nLangeff})  
  obtained for    multiplicative  white   noise, but   the coefficients
 in   these equations are different. 
 Besides,  although
the  effective variables  $\Omega$ and  $Z_2$  used,  
    respectively, for   white noise and colored noise,
   satisfy similar equations,
  they  are   not equivalent [in fact, from Eqs.~(\ref{defZ} and \ref{defY})
 we observe that $Z$ scales as $\Omega^{1/2}$]. 
  Thus,  the   difference between the white  noise and the  colored noise 
  problems  is     embodied 
 in   the successive transformations
 that  relate   $\Omega$   and   $Z_2$  to 
  the   original   energy-angle coordinates.

\subsection{Analytical results}
\label{sec:results}
  
   We now derive the  analytic  expression   for  the P.D.F. of
 the energy of the system  in the  long time  limit and 
   then  calculate  the asymptotic behavior of  physical observables
  such as mean energy, mean position-square and mean
  velocity square. 

  Observing that  the Fokker-Planck equation associated with 
 the  effective Langevin dynamics of $Z_2$,  given by 
  Eq.~(\ref{eq:evolZ2blanc}),  is exactly solvable,  we deduce the
 following expression for the P.D.F.
 \begin{equation}
  P_t(Z_2)  = \frac{2}{\Gamma\left(\frac{n +1}{4(n-1)}\right)}
   \frac{1}{\sqrt{2 \Delta  t   }}      
 \Big(  \frac{  Z_2^2   }{2 \Delta  t } 
  \Big)^{\frac{3-n}{4(n-1)}}   
 \exp\Big( -\frac{  Z_2^2}{ 2 \Delta  t     }\Big)    \, , 
 \label{PDFZ2}
 \end{equation}
 with 
 \begin{equation}
     \Delta    =  (n-1)^2 (\mu_4 - \mu_2^2)  \frac{ {\mathcal D}}{\tau^2} 
  \, .  \label{defDelta}
\end{equation}

     The expression for the  energy $E$ as a function of $Z_2$ can be derived 
 from Eqs.~(\ref{Omega}),~(\ref{defY}),~(\ref{defZ})
 and~(\ref{eq:defZ2}).
 At leading order, we obtain
\begin{equation}
        Z_2 =  2n \, E^{\frac{n-1}{n}}   \, , 
\label{eq:Z2fctE}
\end{equation}
   this relation is valid when $E \gg 1$, {\it i.e.}, in the long time limit.
 We thus deduce  the asymptotic expression,  valid  for 
  $ t \to \infty$,  of the probability distribution function of the energy~: 
\begin{equation}
  P_t(E)  = \frac{2(n-1)}{ n \; \Gamma\left(\frac{n+1}{4(n-1)}\right)}
 \Big( \frac{ 2 n^2 }{  \Delta  t }   \Big)^{\frac{n+1}{4(n-1)}} 
      E^{\frac{1-n}{2n}}
 \exp\Big( -\frac{ 2 n^2 E^{2\frac{n-1}{n}} }{  \Delta  t     }\Big)    \, . 
 \label{PDFEnergy}
 \end{equation}
 If we compare this expression of the  P.D.F. for colored noise
 with the one derived for white noise~(\ref{pdfmult}),
 we observe that the two distribution functions are 
 of the type $ E^\alpha \exp\left( - c E^\beta/ {\mathcal D} t   \right)$,
 with  the characteristic  exponents $\alpha$ and  $\beta$
 and the constant $c$ depending  precisely on nature of the noise.

 The  P.D.F.~(\ref{PDFEnergy}) 
  together with the assumption of uniform angular measure
 in the asymptotic regime leads to analytical expressions
  for  the mean value of any 
 observable of the system.   In particular,
  from Eqs.~(\ref{solnxv2}),~(\ref{equipn2})~and~(\ref{PDFEnergy})
   we  derive the  
  statistical  mean  of the position, velocity and energy of the system
 in the long time limit
\begin{eqnarray}
  \langle E \rangle &=&
  \frac{ \Gamma \left(\frac{3n + 1}{4(n-1)}\right)}
       {\Gamma \left(\frac{n + 1}{4(n-1) }\right)}
\Big(\frac { \Delta t}{2n^2 }\Big)^{\frac{n}{2(n-1)}}
= \frac{ \Gamma \left(\frac{3n + 1}{4(n-1)}\right)}
       {\Gamma \left(\frac{n + 1}{4(n-1) }\right)} 
\Big(   \frac { (n-1)^2 (\mu_4 - \mu_2^2) {\mathcal D} t } {2n^2 \tau^2 }
  \Big)^{\frac{n}{2(n-1)}} 
   \label{moyE} \, , \\
    \langle \dot x^2 \rangle &=&   \frac{2n}{n+1}  \langle E \rangle 
  \label{moyv2} \, , \\
        \langle  x^2 \rangle &=&   \mu_2  \langle E^\frac{1}{n} \rangle 
 = \mu_2 \frac{ \Gamma \left(\frac{n + 3}{4(n-1)}\right)}
       {\Gamma \left(\frac{n + 1}{4(n-1)}\right)}
\Big(\frac { \Delta t}{2n^2 }\Big)^{\frac{1}{2(n-1)}}
  \label{moyx2} \, . 
\end{eqnarray}
 The power-law behaviors  predicted by these  analytical expressions
 are   in accordance  with the qualitative scaling
 relations of Eq.~(\ref{scalingcolor}) 
 deduced   from  heuristic arguments in \cite{philkir2}. 
 Giving some particular values to the parameter $n$,   we obtain~: 
\begin{equation}
\hbox{ for }\,\,\,   n =2, \,\,\,  \langle E  \rangle  = 0.0467 \,
   \Big(   \frac {\mathcal{D} \,t } {\tau^2 } \Big) \, \,\,\,\,   , 
  \,   \langle \dot{x}^2  \rangle  = 0.0623 
   \,\Big(  \frac {\mathcal{D} \,t } {\tau^2 }  \Big) \,\,\,    ,
   \,   \langle x^2  \rangle     =   0.169   \, 
\Big( \frac {\mathcal{D} \,t } {\tau^2 } \Big)^{1/2}     \, , 
 \label{tri}   
\end{equation}
\begin{equation}
 \hbox{ for }\,\,\, n =3, \,\,\,   \langle E  \rangle  = 0.0835 \,
\Big( \frac {\mathcal{D} \,t } {\tau^2 } \Big)^{3/4}   ,
  \,   \langle \dot{x}^2  \rangle  =   0.125 
 \,\Big( \frac {\mathcal{D} \,t } {\tau^2 } \Big)^{3/4}     ,
   \,   \langle x^2  \rangle     =  0.296   \, 
\Big( \frac {\mathcal{D} \,t } {\tau^2 } \Big)^{1/4}     \,  ,
 \label{penta}    
\end{equation}
\begin{equation}
 \hbox{ for }\,\,\,   n =4, \,\,\,  \langle E  \rangle  = 0.0933  \,
\Big( \frac {\mathcal{D} \,t } {\tau^2 } \Big)^{2/3} \,  ,
  \,  \langle \dot{x}^2  \rangle  =  0.149        \,
\Big( \frac {\mathcal{D} \,t } {\tau^2 } \Big)^{2/3}  \,    ,  
   \,   \langle x^2  \rangle  =   0.338     \,
\Big( \frac {\mathcal{D} \,t } {\tau^2 } \Big)^{1/6}      \,  . 
\label{hepta}
\end{equation}

\begin{figure}
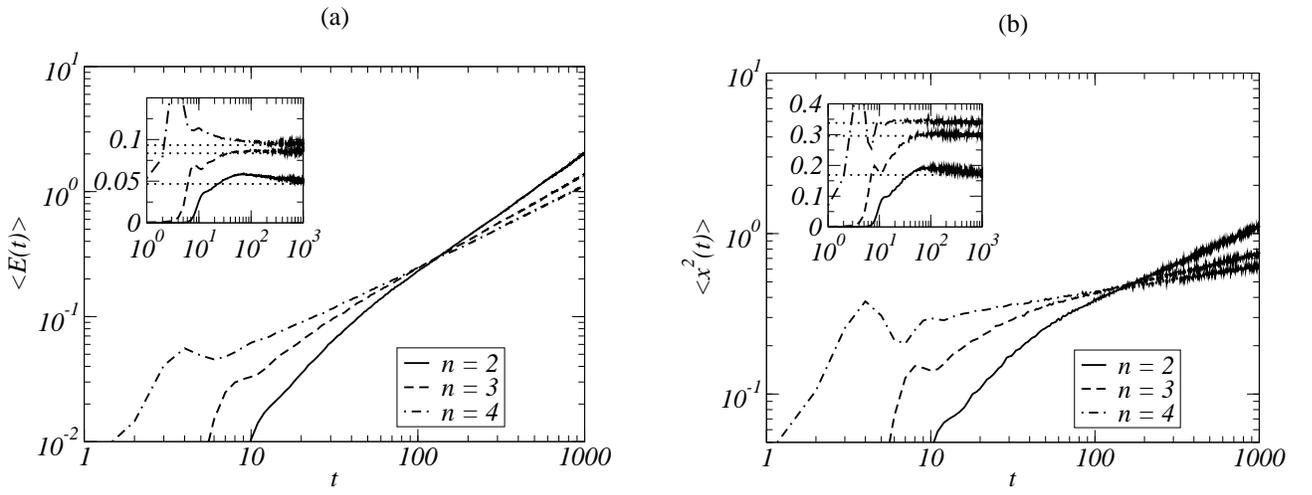

 \hfill
 \includegraphics*[width=0.45\textwidth]{Etau5gam0.eps}
 \hfill
 \includegraphics*[width=0.45\textwidth]{x2tau5gam0.eps}
 \hfill
 \caption{\label{fig:Ex2} Mean value of: $(a)$ the energy 
 and  $(b)$ the square amplitude  of the non-dissipative
 nonlinear oscillator subject  to parametric Ornstein-Uhlenbeck noise, with 
 ${\mathcal D} = 1$ and $\tau =5$. The insets show
the following ratios vs. time $t$:
$(a)$: $\langle E \rangle/(\mathcal{D} \,t/\tau^2)^{n/(2(n-1))}$;
$(b)$: $\langle x^2 \rangle/(\mathcal{D} \,t/\tau^2)^{1/(2(n-1))}$.
The horizontal dotted lines correspond to the numerical prefactors 
given in Eqs.~(\ref{tri},\ref{penta},\ref{hepta}).  }
\end{figure}

  This results  are in excellent agreement with numerical simulations
 (see Figure \ref{fig:Ex2}).

 \subsection{Extension to the case with small dissipation}
 \label{sec:dissipation}
 
 The quantitative analysis described above can be generalized 
 to the study of a stochastic  nonlinear oscillator
 with  very  small  dissipation.  Introducing 
  a linear friction term  with dissipation rate $\gamma$
 in   Eq.~(\ref{dyn2}), we obtain   
\begin{equation}
   \frac{\textrm{d}^2 }{\textrm{d} t^2}x(t)  + 
   \gamma    \frac{\textrm{d}}{\textrm{d} t}x(t)  +  x(t)^{2n-1} 
 = x(t) \, \xi(t)  \,,
 \label{dyndissip}
\end{equation}
 where $ \xi$ is an Ornstein-Uhlenbeck noise. 
 This second order random differential
  equation can be written  
 as a stochastic  system in  the $(\Omega,\phi)$ 
  coordinates \cite{philkir2} 
   \begin{eqnarray}
     \dot \Omega  &=&   -  \gamma  \frac{ n -1}{ (2n)^{\frac{1}{n}} }
  {\mathcal S}_n'(\phi)^2  \Omega  +  
  (n -1) \, {\mathcal S}_n(\phi)
 {\mathcal S}_n'(\phi) \, \xi(t)  
   \,  ,      \label{dissomega} \\
 \dot\phi  &=&  \gamma \frac { {\mathcal S}_n(\phi) {\mathcal S}_n'(\phi)  } 
   { (2n)^{\frac{1}{n}} }
 +  \frac{\Omega}{ (2n)^{\frac{1}{n}}}
 - \frac{{\mathcal S}_n(\phi)^2}{\Omega}  \, \xi(t)  \, .
    \label{dissphi}
   \end{eqnarray}
 Saturation is  attained  when the power injected
 by the random force $\xi(t)$ is balanced by  dissipation.
 The variable  $\Omega$   then  reaches  a finite mean-value
  and fluctuates  around this value.  
 The separation between fast and slow variables is 
 valid only in the limit of a vanishingly small dissipation rate~:
  in this  case,   $\Omega$  saturates to a very large value, 
  the term  $\Omega/(2n)^{\frac{1}{n}} $ is the dominant term  in 
 Eq.~(\ref{dissphi}), and therefore the phase $\phi$ varies
 rapidly with time. We thus  derive an analytic expression 
 for the stationary P.D.F. 
 of the stochastic process~(\ref{dyndissip}) under the hypothesis
 $\gamma \ll 1$, {\it i.e.},  in dimensionless units,   we assume that 
 $\gamma {\mathcal D}^{-1/3}    \ll   1  $  and 
 $\gamma \tau \ll 1$. It is then possible to repeat  all the
 calculations of 
 sections~\ref{sec:transformation},
 \ref{sec:averagingFP}~and~\ref{sec:effectiveLangevin}
  in the presence of a dissipative contribution and to derive the  
  following   effective   white noise  Langevin dynamics 
  for the variable $Z_2$, defined in Eq.~(\ref{eq:defZ2})
 \begin{equation}
   \dot   Z_2  =   - 2\gamma \frac{n -1}{n+1}  Z_2  + 
  \frac{(n-1)(3-n) {\mathcal D}}{ 4 \tau^2} \frac{\mu_4 - \mu_2^2}{Z_2}
  + (n-1)  \frac{ \sqrt{\mu_4 - \mu_2^2}   }{\tau} \eta_1(t)   \, .          
   \label{eq:evolZ2dissip}
  \end{equation}
The stationary Fokker-Planck equation
   corresponding to  this dynamics can be explicitly
 solved and the  following expression 
 for the stationary  P.D.F. of the energy is  obtained  
\begin{equation}
  P_{\rm{stat}}(E)  = 
 \frac{2(n-1)}{ n \; \Gamma\left(\frac{n+1}{4(n-1)}\right)}
 \Big( \frac{ 2 n^2 }{ \tilde  \Delta   }   \Big)^{\frac{n+1}{4(n-1)}} 
      E^{\frac{1-n}{2n}}
 \exp\Big( -\frac{ 2 n^2 E^{2\frac{n-1}{n}} }{  \tilde \Delta}\Big)    \, ,
 \label{PDFstatEnergy}
 \end{equation}
 where $ \tilde  \Delta  $ is given by 
\begin{equation}
  \tilde  \Delta  =  \frac{n+1}{4 \gamma (n-1)}
     \Delta    =  
 (n^2-1) (\mu_4 - \mu_2^2)  \frac{ {\mathcal D}}{4 \gamma \tau^2} 
  \, .  
\label{deftildeDelta}
\end{equation}
 We emphasize that the  P.D.F. given in Eq.~(\ref{PDFstatEnergy})
 is not of  the canonical  Gibbs-Boltzmann  form and therefore does not
 represent a state of  thermodynamic equilibrium. 
 Using this  P.D.F., the moments of
 the energy, position and velocity can be calculated. 
 The analytical  expressions
 of  $\langle E  \rangle$, $\langle \dot x^2  \rangle$  and 
$\langle  x^2  \rangle$  can be deduced  in a formal way  from 
 Eqs.~(\ref{moyE}),~(\ref{moyv2}) and~(\ref{moyx2}) by substituting
  for the time variable $t$ the expression 
\begin{equation}
      \frac{1}{\gamma}\frac{n+1}{4 (n-1)}  \, .
\label{eq:tdissip}
\end{equation}
   This expression defines in fact 
 a    dissipative time scale $t_d$ such  that  
  for $t \ll t_d$ the system behaves as if it were non-dissipative
  and  for $t \gg  t_d$, the system is settled in  a
 non-equilibrium stationary state. For $t \simeq t_d$,
  the  time-dependent  P.D.F.~(\ref{PDFEnergy})
 matches the  stationary  P.D.F.~(\ref{PDFstatEnergy}). We emphasize that
 an important assumption in our calculations is that the  P.D.F.
 is uniform with respect to  the fast  angular variable in the long time limit.  
  This hypothesis   breaks  down when the damping rate $ \gamma$  is high 
   such that the angle and the energy  vary on comparable time scales.  
   The separation between fast and slow variables  is then 
   no more  possible.
   In fact,    when  $\gamma$
   exceeds a critical value (keeping all other parameters fixed)
 the system undergoes a  noise-induced phase transition and  the stationary
 state reduces to   the fixed point $x = \dot x = 0$ \cite{lefever}.
  Although  for white  noise 
 this critical value  is exactly known, an exact   calculation
 of this bifurcation threshold for colored noise 
  remains  open question \cite{crauel}.
  The recursive adiabatic elimination 
  technique developed  here  may perhaps 
  be useful to tackle this   challenging problem.

 \section{Conclusion}

 We   have derived in this work 
 analytical results for the  nonlinear oscillator subject to 
 a  multiplicative colored noise.  
  In the  problem   we have  studied  here, the following 
  features play an essential role~:

 (i) The model's equation   is genuinely of the second order because 
  we   have  considered  the  case   of  zero (or very small)
   damping.  

 (ii) The   dynamics  is  necessarily  nonlinear because we have studied
  the   behavior at large times, for which  the     nonlinearity of
 the confining potential  becomes  relevant.   

 (iii) In the long-time limit, the  finite correlation time $\tau$ of the 
  multiplicative  Ornstein-Uhlenbeck process  is not the
 smallest time scale of the problem (the angular period defines
 a smaller time scale).  Therefore,
 the  noise is genuinely colored and can not be treated
  as a 'quasi-white' even if   $\tau$ is arbitrarily small.

 We emphasize that analytical results for such 
 second-order Langevin equations are scarce.
 Moreover,  in previous works  the above  features 
  are not present   simultaneously. Indeed, different 
  approximations  (for a review see, 
  {\it e.g.}, \cite{hanggirev1,hanggirev2})
  necessarily neglect   one or more aspects  of the problem.  
 For example, in the large damping limit \cite{rahman},
 the inertial term is neglected and the model  is reduced to a first
 order Langevin equation. Short-time behavior is dominated by
 the quadratic part of the confining potential, thus leading to
 a linear stochastic equation. In 
  small correlation-time expansions  that provide 
  effective Fokker-Planck equations \cite{sancho,lindcol}, it is   
  implicitly assumed  that   $\tau$  is  the  smallest time scale; but  
  perturbative expansions  in the vicinity of the white noise limit
    are unable to predict the 
  anomalous  diffusion exponents for  colored noise.
  Specific techniques  such as 
  the Unified Colored Noise Approximation \cite{schimansky,wio} 
  appropriate for
  additive noise problems,  seem to be unsuitable  for  multiplicative
 noise with small dissipation.

       Thanks to a recursive adiabatic averaging   method,
 we have  been able to  perform a mathematical analysis  of  the long
 time behavior of a nonlinear oscillator
  with parametric noise in a non-perturbative
 manner. We have obtained an explicit formula for the P.D.F. in the long
 time limit  which allows us  to calculate the  statistical average
 of any physical observable.  Our  results have been
 successfully compared to numerical simulations; the agreement
 is not only qualitative but also quantitative. In the case
 of a  dissipative system, we have derived an expression 
 for the stationary P.D.F. that compares well with numerical
 simulations in the weak damping regime.

 This study concludes a series of works
 \cite{philkir1,philkir2,philkir4,philkir5}  in which we have 
 generalized the adiabatic averaging technique to derive 
 analytical results for quasi-Hamiltonian nonlinear random  oscillators
 subject to  an additive/multiplicative, white or colored Gaussian noise.

\end{document}